\documentclass[a4paper,12pt]{article}
\usepackage{amssymb,amsmath,mathrsfs}
\usepackage[usenames,dvipsnames]{xcolor}
\usepackage{hyperref}
\usepackage{tensor}
\usepackage{graphicx}
\usepackage{enumitem}
\usepackage[left=1.2in,top=1in,right=1.2in,bottom=1in,headheight=0.8in,foot=0.5in]{geometry}
\usepackage[nodisplayskipstretch]{setspace} 
\usepackage[width=\textwidth]{caption}
\usepackage{pifont}

\usepackage{mathtools}

\newcommand{\nocontentsline}[3]{}
\newcommand{\tocless}[2]{\bgroup\let\addcontentsline=\nocontentsline#1{#2}\egroup}
\hypersetup{
        breaklinks=true,
	colorlinks=true,         
	linkcolor=Black,          
	citecolor=MidnightBlue,
	urlcolor=MidnightBlue            
 }
\usepackage[indentafter]{titlesec}
\titleformat{name=\section}{}{\thetitle.}{0.8em}{\centering\scshape}
\titleformat{name=\subsection}[runin]{}{\thetitle.}{0.5em}{\bfseries}[.]
\titleformat{name=\subsubsection}[runin]{}{\thetitle.}{0.5em}{\itshape}[.]
\titleformat{name=\paragraph,numberless}[runin]{}{}{0em}{}[.]
\titlespacing{\paragraph}{0em}{0em}{0.5em}
\titleformat{name=\subparagraph,numberless}[runin]{}{}{0em}{}[.]
\titlespacing{\subparagraph}{0em}{0em}{0.5em}
\hypersetup{
        breaklinks=true,
	colorlinks=true,         
	linkcolor=Black,          
	citecolor=Blue,
	urlcolor=Blue            
 }
 
\usepackage[style=ext-authoryear-comp,maxcitenames=2,uniquename=false,backend=biber,uniquelist=false,articlein=false, maxbibnames=4]{biblatex}
\DeclareNameAlias{sortname}{last-first}

\AtEveryBibitem{\clearfield{month}}

\title{Promising Stabs in the Dark: Theory Virtues and Pursuit-Worthiness in the Dark Energy Problem}

\author{William J.~Wolf\footnote{Faculty of Philosophy, University of Oxford, UK. william.wolf@philosophy.ox.ac.uk}, \& Patrick M. Duerr\footnote{Carl Friedrich von Weizsäcker Centre for Foundations of Science, University of Tübingen, DE. patrick.duerr@uni-tuebingen.de,  patrick-duerr@gmx.de}}

\date{}

\makeatletter
\let\uppercasenonmath\@gobble

\begin{document}
\setstretch{1.0}
\maketitle

\begin{abstract}

\noindent This paper argues that we ought to conceive of the Dark Energy problem---the question of how to account for observational data, naturally interpreted as accelerated expansion of the universe---as a crisis of underdetermined pursuit-worthiness. Not only are the various approaches to the Dark Energy problem evidentially underdetermined; at present, no compelling reasons single out any of them as more likely to be true than the other. More vexingly for working scientists, none of the approaches stands out as uncontroversially preferable over its rivals in terms of its rationally warranted promise, i.e.~the reasons to further work on, explore, and develop it. We demonstrate this claim by applying a Peircean economic model of pursuit-worthiness in terms of a cognitive cost/benefit estimate---with the instantiation of theory virtues as key indicators of cognitive gains---to the four main Dark Energy proposals (the cosmological constant approach, modified gravity, quintessence, and inhomogeneous cosmologies). According to our analysis, these approaches do not admit of an unambiguous, or uncontroversial, ranking with respect to which ansatz deserves distinguished attention and research efforts. The overall methodological counsel that our analysis underwrites recommends a pragmatic double research strategy forward: to encourage and foster theory pluralism and the search for tests---with the goal of enhancing the testability of the $\Lambda$CDM model and ``testing it to destruction". 


\end{abstract}
\tableofcontents
\setstretch{1.2}

\section{Introduction: the Dark Energy crisis as a problem of triple underdetermination}

\noindent The phenomena subsumed under the umbrella ``Dark Energy" signal 
a major crisis in cosmology. As we'll see, it's an unusual crisis. At their core, the empirical findings that give rise to the Dark Energy mystery boil down to redshift-distance measurements: ``the observed distance to a given redshift $z$ is larger than the one expected from a Friedmann-Lemaître universe with matter only and the locally measured Hubble parameter” \parencite[p.~1]{Durrer:2011gq}. The following syllogism casts the Dark Energy problem into sharper relief:
\begin{itemize}
    \item[(P1)] Assume General Relativity (GR), in its standard form, as our theory of gravity.
    \item[(P2)] Assume the large-scale distribution of matter in the universe to be homogeneous and isotropic.
    \item[(P3)] Assume that the kinds of matter that exist are those described by the Standard Model of particle physics, or Cold Dark Matter.
\end{itemize}
Then, $(P1)\&(P2)\&(P3)\rightarrow (C)$: The Hubble-Lemaître Law's linear relationship between redshift and distance then follows. This, however, clashes with observational data:
 \begin{itemize}
     \item[(O)] The observed distances \textit{exceed} those predicted by the Hubble-Lemaître Law at high redshifts. 
 \end{itemize}  
That is, the speed with which extragalactic systems flee from us seems to accelerate, contrary to the expectations from $(P1)$-$(P3)$: $(O) \rightarrow \neg (C)$. 

Consequently, with $P(1)\& (P2)\&(P3)$ not being consistent with $(O)$, we must relinquish (at least) one of these four premises. That is, we must resort to another theory of gravity, revise our cosmological modelling assumptions about the large-scale structure of the universe, allow for ``exotic" forms of matter, or re-examine and discard the data. 
At first blush, this appears to be a garden-variety instance of the Duhem Problem (see e.g.\ \textcite{ivanova_2021}): logic alone or fixed methodological rules won’t tell us which of the premises to jettison, and what to replace them with.

What makes Dark Energy philosophically and scientifically so exciting---and unsettling!---goes far beyond such a standard Duhemian predicament: resolution of the Dark Energy problem faces a \textit{triple} underdetermination---an exceptionally irksome catch-22.\footnote{The existing philosophical literature (e.g. \textcite{Lahav_Massimi_2014, Butterfield_2014, Jacquart, Koberinski_etal_2023}) correctly identifies underdetermination as the main factor that makes the Dark Energy so tricky. Yet, no attention is paid to the different \textit{kinds} of underdetermination. The third form of underdetermination, in particular---underdetermination of pursuit-worthiness---has been overlooked entirely. This kind of underdetermination also lies at the heart of another crisis in contemporary physics: quantum gravity approaches (\cite{Kiefer,Cabrera_2021}). We thank an anonymous reviewer for pointing this out.} 

First, we have underdetermination of theory by available evidence. Dark Energy phenomenology is compatible with an exuberant multitude of theoretical approaches purporting to account for it \parencite{Wolf_Ferreira_2023, DurrerMaartens, Frieman:2008sn}. This form of underdetermination is familiar from the philosophy of science literature---a recurrent situation in the history of science: the empirical data doesn't uniquely single out a theory. Typically, we hope that this kind of underdetermination is merely transient; one expects future data to ``break” it.

A second kind of underdetermination is more permanent. It concerns situations where future data is \textit{not} expected achieve a resolution. To elaborate: Requisite data would have to ``push the limits of systematic uncertainties''—an ``extremely challenging task'' \parencite[p.~2470]{Perlmutter_2003}. Furthermore, part of the more definitive evidence that would overcome the underdetermination lies in the universe’s ``eschatological future” (arguably long after Humanity has ceased to exist), or at energy scales beyond what seems experimentally accessible \parencite[p.~2]{Ellis}. At present, we therefore have little reason to \textit{expect} data to change this plight. The situation is further complicated by the fact that, in part, it’s not clear what \textit{available} data should count as evidentially relevant: on the one hand, certain coincidences are frequently proclaimed to ``call for an explanation''; on the other hand, one would be hard-pressed to spell out why exactly such coincidences \textit{must} be accounted for.

When confronted with this kind of underdetermination, it's common to invoke some combination of theoretical, methodological, and super-empirical constraints. However, these considerations likewise don’t compellingly whittle down the options. Insofar as the anomalous redshift data is attributed to quantum field theoretic effects, as a manifestation of the quantum vacuum energy, the attempt fails spectacularly (as we’ll discuss in further detail below). Insofar as one decides to modify GR, one plunges into an ocean of possibilities. Most of them are under poor epistemic control, highly speculative, and/or deficient in a persuasive physical motivation.

As if the preceding two forms of underdetermination weren’t disturbing enough, a third kind plagues the Dark Energy problem. The foregoing forms of underdetermination pertain to a paucity of currently available empirical and theoretical constraints for the various proposals that, at least \textit{pro tempore}, doesn’t licence \textit{belief} or \textit{acceptance} of any such option. Belief or acceptance---regarding a theory as well-supported, confirmed, etc.---aren't the only attitudes relevant for scientists, though (see e.g.\ \textcite{mckaughan_dissertation}). Perhaps even more relevant for the practitioner of science are questions of \textit{pursuit}: what theories to invest (time, energy, and other research resources) in, what scientific ideas to work on? This is the third kind of underdetermination of the Dark Energy problem—the sense in which it wreaks such a nerve-wrecking crisis for physicists: no direction of future research seems clearly preferred. This diagnosis will be the present paper's main focus. We’ll defend the claim that no approach to the Dark Energy problem stands out as obviously superior\footnote{This contrasts with another mainstream idea in high-energy cosmology: cosmic inflation. While at least historically it was underdetermined with respect to evidential warrant, i.e. in the context of acceptance, it didn't display underdetermination with respect to pursuit-worthiness. In that regard it was clearly superior to the---empirically adequate!---Hot Big Bang model \textit{sans} inflation and is clearly superior to other alternatives (especially) due to its fertility in displaying instances of problem and predictive novelty \parencite{McCoy2015-MCCDIS, Wolf_Duerr_2023, Wolf_2023, Wolf_Karim_2023}.} in terms of pursuit-worthiness; solutions to the Dark Energy problem are underdetermined with respect to pursuit.\footnote{Note how this underdetermination of pursuit-worthiness \textit{doesn’t} apply to the Dark Matter problem: the existence of cold dark matter is sufficiently epistemically warranted thanks to independent lines of evidence and its coherence with respect to rest of established physics corpus (in contrast with competing theories like MOND and relativistic extensions; see \textcite{DuerrWolf_2023} for details). This doesn't mean, of course, that the Dark Matter problem is free from other forms of underdetermination—especially those concerning the \textit{physical nature} of Dark Matter \parencite{Martens_2022}.}

Given this underdetermination of pursuit-worthiness, we plea for modest conservatism as the methodological recommendation for the way forward---as a \textit{useful heuristic}\footnote{cf.~\textcite{DeBaerdemaeker_2020} for an analogous case for Dark Matter}: on the one hand, to pursue the conservative approach, the $\Lambda$CDM model, especially with the goal of ever stricter tests; and on the other hand---the complementary modesty (or anti-dogmatism/liberalism) of the recommended conservatism---to embolden the foraging for unconventional answers to the Dark Energy probem.

We'll proceed as follows. \textbf{§2} will elucidate the Dark Energy problem in greater detail, and sketch the gist of the principal options for a solution. To tackle the question of their pursuit-worthiness, we'll also expound our philosophical rationale for adjudicating whether a theory merits further research. \textbf{§3} will apply them to the main Dark Energy proposals to assess their pursuit-worthiness. We conclude in \textbf{§4}.

\section{Accelerated cosmic expansion: facts, ideas, and the question of pursuit}

\noindent Here, we'll first (\textbf{§2.1}) present an exposition of the empirical side of the Dark Energy problem. \textbf{§2.2} will review the basics of the four main approaches to Dark Energy, and how they are supposed to account for the Dark Energy phenomenology. In \textbf{§2.3}, we'll outline the rationale for justifying pursuit that we'll adopt, a Peirce-inspired cognitive cost/benefit analysis. 

\subsection{Dark Energy phenomenology}

The starting point of modern cosmology consists in applying GR to the universe at large \parencite{Baumann_2022, weinberg_2008}. Cosmological models are thus solutions to the Einstein field equations 
\begin{equation}\label{EFE}
R_{\mu\nu}-\frac{1}{2}Rg_{\mu\nu} = \kappa T_{\mu\nu},
\end{equation}
where $\kappa := 8\pi G/c^4$. They express a relationship between $R_{\mu\nu}$, the Ricci curvature, associated with $g_{\mu\nu}$, the spacetime metric, and the stress-energy tensor of matter, $T_{\mu\nu}$. The latter encodes modelling assumptions about the spatial distribution of matter on cosmic scales. Standard cosmology treats cosmic matter---galaxies and galaxy clusters---as a perfect (i.e.\ non-viscous and shearfree) fluid, or collisionless ``dust” particles, characterised by its rest frame mass density $\rho$ and isotropic pressure $p$. 
The associated stress-energy tensor takes the form (with the fluid ``molecules'" four-velocity $u_\mu)$:
\begin{equation}\label{perf-fluid}
    T_{\mu\nu} = (\rho + p) u_\mu u_\nu + p g_{\mu\nu}.
\end{equation}
Solutions to the Einstein Equations yield the so-called Friedmann-Robertson-Lemaître-Walker (FLRW) cosmological models. They are given by the FLRW spacetime\footnote{Conveniently written in terms of the line-element $ds^2= g_{\mu \nu}dx^\mu dx^\nu$.}:
\begin{equation}
    ds^2 = -c^2dt^2 + a(t)^2 \left[ \frac{dr^2}{1 - k r^2} + r^2 (d\theta^2 + \sin^2\theta d\phi^2) \right].
\end{equation}
Here, $k$ denotes the \textit{spatial} curvature\footnote{That is, the curvature of the three-dimensional hypersurfaces corresponding to simultaneity time-slices $t:=x^0=const.$}, and $a(t)$ the so-called scale factor. It determines physical lengths. 
That it varies in time is precisely the meaning behind the statement that our universe expands.
The dynamics of $a(t)$ are governed by the so-called Friedmann Equations, a pair of coupled ordinary differential equations which, under the above assumptions, the Einstein Equations reduce to: 
\begin{equation}\label{Friedmann1}
    H^2 \equiv \left(\frac{\dot{a}}{a}\right)^2 = \frac{8 \pi G}{3}\rho - \frac{k^2}{a^2},
\end{equation}
\begin{equation}\label{acceleration}
    \dot{H}+ H^2 \equiv \frac{\ddot{a}}{a} = -\frac{4\pi G}{3}\left(\rho + 3\frac{p}{c^2}\right).
\end{equation}
Here $H(t):= \dot{a}(t)/a(t)$ represents the so-called Hubble ``parameter”. It quantifies the universe's rate of expansion. 

A wealth of data indicates that we inhabit a spatially flat ($k \approx 0$) universe (see e.g.\ \textcite[Ch.9]{Peebles2020}). At the same time, the observed mass density $\rho$ (including all forms of $\rho$ and $p$ given by the contributions from baryonic matter, Dark Matter, and radiation) is too low for a flat universe. If one allows for unknown sources of---accordingly monikered ``\textit{Dark}”---energy, likewise uniformly distributed, with pressure $\rho_\text{DE}$ and $p_\text{DE}$, to contribute to the universe’s total energy budget, 
the data can be consistently accommodated. The total energy density, then, is roughly parsed into 30\% contributions from (mostly Dark) matter and 70\% Dark Energy. For $k=0$, the measured $\rho$ and $p$, and inferred $\rho_\text{DE}$ and $p_\text{DE}$, the Friedmann Equations imply an accelerated increase in the scale factor; hence, the phenomenology encapsulated in the data is naturally interpreted as the universe’s accelerated expansion.

We primarily infer Dark Energy from cosmic redshift/distance measurements. Redshift measurements are obtained through the familiar analysis of spectral shifts. Distance measurements are procured a little more obliquely. One uses two main notions of distance: luminosity distance and angular diameter distance (see e.g.~\textcite[Ch.~2]{Baumann_2022}). The luminosity distance relates the intrinsic luminosity of an object to its observed brightness, while angular diameter distance refers to the ratio between an object's physical size and its observed size. Both of these distances are ways of translating intrinsic properties of the objects of interest into the quantities that we perceive as observers. These observer-dependent distances are functions of the cosmological co-moving distance. It describes cosmological distances across different epochs, as it remains constant over time and scales with the expansion of the universe. The expression for co-moving distance is given by:
\begin{equation}\label{comoving distance}
D_C (z) = \int_0^z \frac{d z^{\prime}}{H\left(z^{\prime}\right)},
\end{equation}
with redshift $z$ and the Hubble parameter $H (z)$ as a function of redshift. 
Luminosity and angular diameter distance are related to the co-moving distance via
\begin{equation}
D_L = (1+z) D_C; \quad D_A = D_C/(1+z). 
\end{equation}
The factors $1+z$ and $(1+z)^{-1}$ account for how, in an expanding universe, the redshift of light impacts observations. The luminosity distance is larger than the co-moving distance: due to the stretching of space, objects appear dimmer and thus farther away. By the same token, angular diameter distance, reflecting an object’s apparent size in the sky, becomes smaller.

Astronomers measure the luminosity distance $D_L(z)$ to Type Ia supernovae. Thanks to their known luminosity profiles, the latter serve as standard candles. If we \textit{only} assume radiation and matter as sources of energy density in our cosmological modelling, the FLRW models don’t match the data: given the redshifts observed for the supernovae, their predicted distances $D_L(z)$ aren’t large enough. Supplementing the universe's energy density with a significant Dark Energy component brings the expected and observed $D_L (z)$ into agreement.
Similarly, measurements of the Cosmic Microwave Background (CMB) and Baryonic Acoustic Oscillations (BAOs)\footnote{BAOs are periodic fluctuations in the density of baryonic matter in the early universe, which leave an imprint on CMB and the distribution of galaxies observed today.} are sensitive to angular diameter distance measurements. They yield ``missing'' energy contributions consistent with the results of the foregoing supernovae measurements: the universe contains $\sim 70\%$ Dark Energy \parencite{Planck:2015bue}.

For our presentation of Dark Energy models in \textbf{§2.2}, it will be instructive to further explicate the case of the supernovae measurements of $D_L (z)$, and how they imply the presence of Dark Energy. As per (\ref{comoving distance}), luminosity distance is a function of $H(z)$. In turn, we can write $H(z)$:
\begin{equation}\label{Hubble}
    \frac{H(z)}{H_0} = \left[\Omega_{\mathrm{rad}} a^{-4}+\Omega_{\mathrm{M}} a^{-3}+\Omega_\text {DE}a^0\right]^{1/2} = \left[\Omega_{\mathrm{rad}} (1+z)^{4}+\Omega_{\mathrm{M}} (1+z)^{3}+\Omega_\text {DE}\right]^{1/2},
\end{equation}
with the Hubble parameter’s current value $H_0 \coloneqq H(t)|_\text{now}$. Here, we exploited the fact that redshift and scale factor are related by $(1+z) = a^{-1}$, and  expressed the energy components as fractions of the so-called critical density $\rho_c \coloneqq 3H^2 / 8 \pi G$. In a flat universe---such as ours—the total energy coincides with it: $\Omega \coloneqq 8\pi G \rho_c/(3H^2) = \Omega_{\mathrm{rad}} + \Omega_{\mathrm{M}} + \Omega_\text {DE}= 1$; where these are the energy densities of radiation ($\Omega_{\mathrm{rad}}$), matter ($\Omega_{\mathrm{M}}$), and an unknown one, $\Omega_\text{DE}$, from Dark Energy. For the sake of simplicity, let’s for now assume that $\Omega_\text{DE}$ doesn’t dilute with expansion. Suppose, that is, it doesn’t depend on $z$ (or, equivalently, $a$). In a universe with a $\Omega_\text{DE}$ component, $\Omega_\text {DE}$ will quickly dominate the energy budget as the energy densities for radiation and matter rapidly thin out due to their dependence on $z$.

This will enhance $H(z)$---\textit{accelerating} the universe’s expansion---and any distance measurements. In practice, one directly measures $D_L(z)$ (and then determines $H(z)$ through these observations); the properties of the various energy components are inferred by fitting these measurements of $D_L(z)$ to theoretical models of $H(z)$.\footnote{See e.g.~\textcite{Wolf_Ferreira_2023, DESI_Wolf:2024eph} for an example of how this is done with quintessence models. The equation of state is typically parameterised as a Taylor expansion $w(a)\simeq w_0 + w_a (1-a)$. Here $w_0$ represents the value of the equation of state now, while $w_a$ captures its time variation. One determines these $w_0$, $w_a$ observables by fitting the model of $H(z)$ using this parameterisation to measurements of $H(z)$.} Supernovae observations indicate that $H(z)$ is significantly larger than one would expect in a universe dominated by matter. The best fits for $\Omega_{\mathrm{rad}}$, $\Omega_{\mathrm{M}}$  and $\Omega_\text{DE}$ correspond to models for a universe dominated by a constant energy component $\Omega_\text{DE}$. 

If the data are reliable and our assumptions---in particular, the applicability of GR on cosmological scales, and the uniform large-scale distribution of matter---are correct, we are searching for an energy component with an equation of state (the ratio between pressure $p$ and energy density $\rho$) close to $w _\text{DE}:= p_\text{DE}/\rho_\text{DE} \simeq -1$.\footnote{This is because $\rho \propto a^{-3(1+w)}$ and $w \simeq -1$ corresponds to $\rho \propto a^{0}$, i.e.\ an energy component that doesn’t dilute with expansion.} A Dark Energy proposal, in short, saves the Dark Energy phenomena, when it reproduces an effective equation of state approximately $-1$. Regarding the current status of the data, there is still a significant amount of open parameter space in our phenomenological characterisation of $w$ (see Fig.~4 in \textcite{Planck:2015bue} or Fig.~6 in \textcite{DESI:2024mwx}). Until the most recent DESI results in \textcite{DESI:2024mwx} from earlier this year, all CMB, SNe, and BAO data had been consistent with, and indeed favoured, a cosmological constant where $w=-1$ (to be discussed shortly). But the latest tranche of data has pulled away from a cosmological constant at $\sim 3\sigma$ and may be hinting at a dynamical equation of state.\footnote{These new results have generated a large amount of debate and discussion from both the data and theory sides since their release. See e.g.~\textcite{DESI_Cortes:2024lgw, DESI_Dinda:2024ktd, DESI_Tada:2024znt, DESI_Wolf:2024eph, DESI_Ye:2024ywg, DESI_Wolf:2024stt, DESI_:2024kob} for a representative sample of recent analysis.} These are exciting times for dark energy research, and more data will hopefully give us more clarity on the \textit{nature} of this, so-far merely phenomenologically characterised, form of energy.

\subsection{Key aspects of main Dark Energy models}

Here, we'll provide a brief overview of the main Dark Energy proposals, preparing our more detailed appraisal of their pursuit-worthiness in \textbf{§3}. Our aim for now is to introduce their basic physics, and render transparent how they seek to account for Dark Energy phenomenology.

\subsubsection{Cosmological Constant}

A cosmological constant $\Lambda$ can be straightforwardly inserted into GR's field equations (\ref{EFE}) via, 
\begin{equation}\label{CCfieldeq}
R_{\mu\nu}-\frac{1}{2}Rg_{\mu\nu}+\Lambda g_{\mu\nu} = \kappa T_{\mu\nu},
\end{equation}
or, equivalently, in the action
\begin{equation}\label{CCaction}
    S_\text{EH} = \int d^4 x\sqrt{-g} \left[\frac{1}{2\kappa} (R - 2\Lambda) + \mathcal{L}_m\right].
\end{equation}
Here $g$ is the determinant of the metric, $R$ is the Ricci curvature scalar, $\mathcal{L}_m$ is the matter Lagrangian (see e.g.\ \textcite[Ch.~19]{HobsonEfsthLasen}).

The effect of $\Lambda$ in (\ref{CCfieldeq}) can be formally written 
in terms of GR's standard form, (\ref{EFE}), with an extra $\Lambda$-dependent energy-stress tensor (of the perfect fluid type (\ref{perf-fluid}), with constant density $\rho_\Lambda := \Lambda / {c^2}$ and \textit{negative} pressure $p_\Lambda := -\rho_\Lambda c^2 =- \Lambda$ )
\begin{equation}
T^{(\Lambda)} _{\mu \nu} := - \Lambda g_{\mu \nu}.
\end{equation}
This leads to \textit{exactly} the requisite equation of state:
\begin{equation}
    w_{\Lambda} \equiv -1.	
\end{equation}
The criteria laid out in \textbf{§2.1} for Dark Energy phenomenology are thus clearly satisfied.

\subsubsection{Modified Gravity}

There are a large variety of ways one can go about modifying gravity, including, but not limited to, scalar-tensor theories, vector-tensor theories, bi-metric theories, higher order theories, etc. \parencite{Clifton_etal}. We'll focus on so-called $f(R)$ theories of gravity and treat them as a representative for modified gravity theories. While $f(R)$ theories are technically within the class of higher order theories (as they include higher order curvature invariants), they also share much in common with scalar-tensor theories (arguably the other most extensively studied class of modified gravity theories).

$f(R)$ Gravity modifies Einstein’s GR by changing the latter’s gravitational dynamics (i.e.\ the dynamics for the metric field), whilst retaining the way gravity couples to ordinary (non-gravitational) matter \parencite{Sotiriou_Faraoni_2010}. In particular, $f(R)$ Gravity starts from GR’s Einstein-Hilbert action with the metric’s Ricci scalar $R$ as the action’s Lagrangian and generalises this density by replacing the Ricci scalar with a general function of it, $R\rightarrow f(R)$:
\begin{equation}
    S_{EH} = \int d^4 x\sqrt{-g} \left[\frac{1}{2\kappa}f(R) + \mathcal{L}_m\right].
\end{equation}

The ``effective'' equation of state $w_{f(R)}$ for an $f(R)$ Gravity cosmological model can be written as: 
\begin{equation}\label{f(r)w}
w_{f(R)} \equiv \frac{P_{f(R)}}{\rho_{f(R)}}=-1+\frac{2 \ddot{F}-2 H \dot{F}-4 \dot{H}\left(F_0-F\right)}{(F R-f(R))-6 H \dot{F}+6 H^2\left(F_0-F\right)},
\end{equation}
where $F=df(R)/dR$, $H$ represents the Hubble parameter, and $F_0$ represents the present-day value of $F$ (see \textcite{Amendola:2006we} for further details). To secure phenomenological viability (including e.g., cosmological, astrophysical, and solar system constraints) one chooses suitable functions $f$.

\subsubsection{Quintessence}

The guiding thought behind quintessence models is to allow for a dynamically evolving alternative to the cosmological constant \parencite{Peebles_Ratra_1988, Caldwell_etal_1998}. Quintessence is typically viewed as an exotic form of matter. It's characterised by a dynamical scalar field $\varphi$ with a canonical kinetic term and a potential $V(\varphi)$ and couples minimally to gravity (but \textit{not} to other forms of matter). 

The action is accordingly given by:
\begin{equation}
S=\int d^4 x \sqrt{-g}\left[\frac{1}{2\kappa} R-\frac{1}{2} g^{\mu \nu} \partial_\mu \varphi \partial_\nu \varphi-V(\varphi) + \mathcal{L}_m\right].
\end{equation} 
For the potential---so-far a free function---a suitable form is chosen so as to reproduce the desired late-time cosmic acceleration (as well as to conform to other physical constraints, furnished by e.g.\ nucleosynthesis and cosmic structure, which are sensitive to such a field). Most crucial in this regard is that during late times the potential is dominant; the quintessence field’s equation of state, $w_{\varphi}$, then approximates that associated with the cosmological constant:
\begin{equation}\label{DEeq}
w_{\varphi} =\frac{\dot{\varphi}^2 / 2-V(\varphi)}{\dot{\varphi}^2 / 2+V(\varphi)}\approx -1,
\end{equation}

Primarily two types of potentials are considered: (i) so-called thawing models, and (ii) so-called freezing models. Thawing models ``thaw" away from $w\simeq-1$ at late times as they evolve, while freezing models ``freeze" at late times as they evolve towards $w\simeq-1$  \parencite{Tsujikawa_2013, Caldwell_Linder_2005}.

\subsubsection{Inhomogeneous Cosmologies}

Up to this point, we have tacitly granted a basic assumption behind FLRW models: that the universe's large-sale structure is homogeneously and isotropically distributed. Should it  be incorrect, would we still be able to account for Dark Energy phenomenology? Cosmologists have explored this in so-called backreaction \parencite{Ellis:2011hk, Buchert_2011} and void models \parencite{Mustapha:1997fjz, Celerier:1999hp, Celerier:2007jc}:
\begin{enumerate}
    \item \textbf{Backreaction models} reject the standard assumption that linear perturbation theory is a sufficient approximation to describe late-time cosmic dynamics. That is, they recover FLRW dynamics at lowest order, but allow for non-linear corrections that reproduce the dynamical effects of accelerated expansion.
    The models seek to explain a \textit{real} cosmic acceleration as a cumulative effect of these neglected \textit{small-scale} inhomogeneities. 
    \item \textbf{Void models} reject the FLRW assumption of homogeneity while retaining the assumption of isotropy. Large inhomogeneities, such as cosmic voids, can generate different local expansion rates, depending on the matter density in one's local environment.  \textit{No true} global cosmic acceleration exists; only an apparent acceleration is induced by \textit{large-scale} inhomogeneities.
\end{enumerate}

Backreaction models pivot on the recognition that owing to GR’s non-linear mathematical structure, averaging an \textit{in}homogeneous solution to the field equations (\ref{EFE}) doesn’t necessarily result again in another solution. One can’t \textit{naively} smooth out inhomogeneities in cosmological models, and proceed to work with an idealised homogeneous, coarse-grained proxy. Intriguingly, by incorporating a kinematical backreaction term (quantifying non-linear effects) adherents of the backreaction programme succeed in producing equations formally very similar to the Friedmann Equations \textit{with} a Dark Energy contribution (see \textcite{Buchert_2011} for details). Here, the cosmic parameters---energy density, spatial curvature, etc.---are averaged over a chosen domain. Distance measurements can be fitted to the resulting ``effective” FLRW metric without exotic matter or modifying gravity \parencite{Desgrange_etal_2019}.  

The starting point of void models is the Lema\^{\i}tre-Tolman-Bondi (LTB) metric, a spherically symmetric, but radially inhomogeneous cosmological model:
\begin{equation}
    ds^2 =  -dt^2 + \frac{(R'(r,t))^2}{1 + 2E(r)} dr^2 + R(r,t)^2 (d\theta^2 + \sin^2\theta d\phi^2).
\end{equation}
Here, $R(r,t)$ (with its radial derivative $R'(r,t) := \partial R(r,t) / \partial r$), representing a generalised, time-/radius-dependent scale factor, and $E(r)$, representing a generalised, time-radius-dependent spatial curvature, are free functions, alongside an additional freedom in the mass distribution, via $m(r)$. This defines a distance function that can be compared to distance-related observations. Intuitively, void models seek to account for apparent cosmic acceleration by locating us in a large cosmic void. Such a region expands faster than the rest of the universe, having comparatively less matter (and thus less gravitational attraction to slow expansion). Note that, in void models, there is no global cosmic acceleration but only \textit{local} differences in expansion rate, driven by local differences in matter density.

\subsection{The Dark Energy problem as underdetermination of pursuit-worthiness}

The Dark Energy problem oughtn't be conceived of as a crisis in the accustomed sense---that is, the sense inspired by \textcite{Kuhn1962-KUHTSO-3}: it would be misleading to characterise it as the frenzied explorations of motley alternatives to the ruling paradigm, the standard $\Lambda$CDM model of cosmology, \textit{triggered} by an empirical or theoretical anomaly that defies the existing $\Lambda$CDM framework (see \textcite{Planck:2015fie, Perivolaropoulos:2021jda, Bull:2011wi}).\footnote{Needless to say, anomalies \textit{do} haunt the $\Lambda$CDM model (see e.g.\ \textcite{Lopez-Corredoira:2017rqn}). Amongst them, some Dark Matter-related galactic phenomena ( \textcite{McGaugh}), or the Hubble tension (the divergence of the Hubble constant’s value, as determined by two different methods, see e.g.\ \textcite{Smeenk2022-SMETWH-2}) stand out. Notwithstanding their importance in their own right, however, these challenges to the cosmological standard model, are typically \textit{not} meant by ``the Dark Energy crisis”.} Instead, it would be more astute to see it the other way round: the Dark Energy crisis consists in a proliferation of simultaneously pursued approaches to account for the pertinent phenomena, a plurality of competing ideas---in the \textit{hopes} of hitting on a unimpeachable anomaly for the standard model.  

Vis-à-vis the multitude of options, which approach should be adopted? Absent conclusive evidence which singles out one over the other, one might be inclined to respond: none! Such a lackluster verdict would---fortunately---be a little rash: to adopt an approach needn’t be exclusively construed in terms of evidential support, that is, in terms of reasons to regard it as epistemically warranted (or, for those with realist sympathies: an approximately true description reality). Instead, one can---and scientists indeed routinely do---adopt an attitude that suspends, for the time being, judgements of warrant and rather asks: does the approach merit further study and elaboration? The Dark Energy problem concerns an area of cutting-edge---and not-yet-epistemically-warranted!---research; hence this second question seems especially apt.

In other words, heeding a distinction initially drawn by \textcite[p.~108]{Laudan1977-LAUPAI}\footnote{See also \textcite[p.~111]{Laudan1996-LAUBPA-3}, \textcite{Seselja2012} or \textcite{Barseghyan_Shaw_2017}.}, we should ask: which approach to Dark Energy should we \textit{pursue}---invest cognitive resources (and time and manpower) in its  further exploration and development (perhaps \textit{until} an approach is amenable to a more conclusive evaluation in terms of acceptance on the basis of evidence or support)?

For a principled answer, we must flesh out a rationale for justifying (cognitive\footnote{We’ll limit ourselves to cognitive criteria for pursuit-worthiness, i.e.\ reasons anchored in cognitive or epistemic, science-inherent promise---rather than technological or more extraneous benefits (such as idiosyncratic curiosity or a wealthy patron, sponsoring particular research projects).}) pursuit: by which criteria to rationally decide whether an approach merits further pursuit?
As our subsequent philosophical working hypothesis, we’ll adopt Peirce’s ``economic” account of pursuit-worthiness (\textcite{Nyrup2015-NYRHER, McKaughan_2008}; cf.~\textcite{DouglasValues, Seselja_2014, Lichtenstein_2021, Fleisher_2022}), a utility estimate, or cost/benefit analysis of the anticipated cognitive gains hoped for when investing in the idea-to-be-pursued: a theoretical approach to a scientific problem counts as pursuit-worthy if its (plausibly) expected epistemic-cognitive benefits (plausibly) outweigh its (plausibly) expected costs.\footnote{The fertility of the economic account of pursuit-worthiness has been demonstrated in several case studies (e.g.~\textcite{Wolf_Duerr_2023, SeseljaWeber, Schindler_2014, RuizdeOlano_2023, Cabrera_2021, Fischer2024-FISTPO-31}). In particular, we take these works to show that appraising theories in terms of a theory virtue-based cognitive cost/benefit analysis is a powerful tool that is capable of nuanced and substantive verdicts.} 

Here, costs are to be understood as expenditures of resources, first and foremost research time and efforts. In a pursuit-worthy idea, they are supposed to be offset by potential cognitive pay-off. Ideally, this means an empirically successful and cognitively valuable theory. By the latter we mean a theory that instantiates certain prima facie attractive features, so-called theory virtues, such as simplicity, coherence, explanatory or unificatory power \parencite{Keas2018-KEASTT-2, IvanovaForth, mcmullin2014virtues, Kuhn1977-KUHOVJ}. Theories or frameworks exhibiting them hold immense cognitive promise; they pledge such desirable accomplishments with respect to the aims or values of science, first and foremost ``to find satisfactory explanations of whatever strikes us as being in need of explanation'' \parencite[p.~132]{Popper1983-POPRAT}. But also non-optimal benefits should be factored into the utility assessment: they may include important spin-off lessons for, say, calculational techniques or qualitative insights into the general capacities of a modelling approach, likely to be gained from pursuing an idea (whatever the ultimate verdict on its viability). Note that some theory virtues—those typically regarded as pragmatic (rather than truth-conducive), simplicity in particular---may not only indicate potential cognitive gain; they may also lower cognitive costs (by e.g.~making an approach easier to handle). In other words, justifying pursuit is ``a decision theoretic problem of how to optimize the epistemic output of science'' \parencite[p.~753]{Nyrup2015-NYRHER}.

Having now assembled the philosophical tools, our next task is to apply them to the Dark Energy approaches of \textbf{§2.2}: what guidance does the economic account of pursuit-worthiness offer?  

\section{Analysis: the pursuit-worthiness of Dark Energy proposals}

\noindent This section will inspect the four main proposals for the Dark Energy problem: the cosmological constant approach (\textbf{§3.1}), the modified gravity approach (\textbf{§3.2}), quintessence models (\textbf{§3.3}), and inhomogeneous cosmologies (\textbf{§3.4}). We'll gauge their pursuit-worthiness in terms of the theory virtues and vices they instantiate. Our analysis will foreground what makes the Dark Energy problem so bewildering: a deadlock of pursuit-worthiness. 

\subsection{The cosmological constant approach}

We begin with the most common and orthodox approach to DE: the cosmological constant. 

\subsubsection{Coherence and simplicity}
Introducing a cosmological constant $\Lambda$ 
isn't \textit{merely} the most natural option to accommodate Dark Energy phenomenology within GR. We may plausibly view it as \textit{the} general-relativistic Dark Energy account tout court. The two standard approaches to GR---the heuristic/motivational paths leading to it, one based on formal desiderata (the so-called Lovelock Theorem, see e.g.\ \textcite[Sect.~2.4]{Clifton_etal}) and the other on physical principles (reconstructed in \textcite[Ch.~7.1]{Weinberg:1972kfs})---both determine the field equations only up to the form of Eq.~\eqref{CCfieldeq} or Eq.~\eqref{CCaction}, which includes a cosmological constant. In this sense, one shouldn't view the cosmological constant as an \textit{addition} to, or modification of, GR. Rather, it forms an essential part of it: a free parameter, encoding the inherent tendency of (uniform) spacetime to exponentially expand. As such, it portends nothing mysterious. Within the cosmological constant approach, $\Lambda$ merely denotes a constant of nature, representing space's ``elasticity", in the general-relativistic law of gravity, akin in its status to $c$ or $G$ \parencite{Bianchi:2010uw}. This non-adhocness of the cosmological constant is undoubtedly one of its greatest assets \parencite[Sect.~4.5.1]{DuerrWolf_2023}. 

One may perhaps resent the presence of any free parameter as a source of ad-hocness (for instance, along the lines of \textcite{Hitchcock2004-HITPVA} or \textcite{Worrall2014-WORPAA}): whatever its particular value, such a contingent element somewhat blights the theory. Note that in this regard, it’s irrelevant whether its numerical value is zero: not having a free parameter \textit{isn't} the same as having a free parameter that takes the value zero!\footnote{Even if other Dark Energy approaches postulate a vanishing cosmological constant, this particular choice requires a justification. Typically, none is one given, though. We thank an anonymous reviewer for pressing us on this!} For greater coherence\footnote{By ``coherence” we mean in the following a theory's organic structure, the existence of tight links amongst its ingredients, and amongst the theory and other theories (see \textcite{BonJour1985-BONTSO-4,Seselja_2014} for details). In a coherent theory, its parts ``hang together”---both amongst themselves (``internal coherence”), as well as with respect to other parts of our knowledge, or at least ideas deemed pursuit-worthy  (``external coherence”). Lack of coherence, we maintain, detracts from a theory’s pursuit-worthiness: it taints its promised epistemic gains. Egregious forms of deficient coherence, or of incoherence, may even be regarded as tantamount to epistemic losses.}, it’s desirable to eschew free parameters altogether.

The cosmological constant's \textit{specific} numerical value is typically perceived as its most disconcerting aspect: ``(a)mong cosmologists, recent thinking about evolution of the parameters of physics is dominated by (its)] curious value" \parencite[p.49]{Peebles2020}: almost---but not exactly---zero, the current value being $\Lambda = (2.846 \pm 0.076) \times 10^{-122}\text{m}^2_{\text{pl}}$ \parencite{Planck:2018vyg}. One covets an explanation: given the infinity of possibilities, the value that yields the best fit for the data appears \textit{unexpected} and oddly fine-tuned. The impression of fine-tuning is also aggravated by the qualitative cosmological ramifications of a small change in $\Lambda$. It can affect the long-term destiny of the universe: whether it re-collapses, or undergoes decelerated or accelerated expansion (see e.g.\ \textcite[Sect.~9]{Frieman:2008sn}). Hence, one may rebuke the cosmological constant---notwithstanding the solid motivation for its existence---as ad-hoc, as far as its value is concerned: it looks ``suspiciously" close to zero to be shrugged off as a coincidence; translating it into a mass/energy scale the cosmological constant seems to stray far from other characteristic mass/energy scales. But such a verdict, which would \textit{dis}courage pursuit, hinges on controversial premises (cf.\ \textcite{Hossenfelder:2018ikr}): how trust-worthy are our expectations? What’s their physical, or even empirical, basis? Do they go beyond subjective hunches\footnote{Usually the issue is phrased in terms of a suitable biased probability distribution over values of the cosmological constant, see e.g.\ \textcite[p.~32]{Carroll:2000fy}. The worry urged above resurfaces: what’s the physical---or, more generally, epistemic---basis of such alleged probability distributions (cf.\ \textcite{Norton2018})} or, likewise epistemically dubious aesthetic preferences? 

\subsubsection{External coherence}

At any rate, a free parameter strikes many as a rebarbative form of ad-hocness. Might recourse to \textit{other} theories supply $\Lambda$'s numerical value? This would kill two birds with one stone: one would thereby mitigate its ad-hocness (qua free parameter), and boost the approach’s external coherence (i.e.\ its meshing and inferential links with the rest of our background corpus of beliefs and theories).

The popularity of that hope therefore comes as no surprise (see \textcite{Kragh:2014jaa} for the history of that quest). It inspired major research efforts. Regrettably, to-date, they haven’t borne fruit. 

In the late 60s, working within a larger tradition in Soviet physics that sought an enticing bridge between high-energy particle physics and cosmology, Zel’dovich had re-examined Lemaître’s earlier (but underdeveloped) idea of regarding $\Lambda$ as the vacuum energy. He now associated it with the zero-point fluctuations in quantum field theory (of whose reality most physicists had been convinced after the corroboration of the Casimir effect in the late 1950s). On the one hand he showed that such fluctuations reproduce the mathematical \textit{form} of the cosmological constant---a tantalising result. On the other hand, however, the numerical value “greatly exceeded any reasonable observational bound on the value of the cosmological constant” \parencite[p.~63]{Kragh:2014jaa}. Others quickly picked up Zel’dovich’s baton. To no avail: the discrepancy between viable options for $\Lambda$ and quantum-field theoretic vacuum energy remained (and remains) gigantic. The efforts culminated in \textcite{Weinberg:1988cp}'s no-go theorem: ``under very general conditions, [the interpretation of $\Lambda$ as the quantum-field theoretic vacuum fluctuations] (can) not account for a small or vanishing cosmological constant without an extremely high degree of fine-tuning” \parencite[p.~102]{ORaifeartaigh:2017yby}. Such a cure for the ad-hocness of a free parameter sounds worse than the disease: why trade the moderate ad-hocness of a (well-motivated) free parameter within a classical/phenomenological theory for severe structural ad-hocness in the form of inordinately contrived modelling assumptions for quantum-field theoretical dynamics?  
The attempt to interpret $\Lambda$ as the vacuum energy has subsequently been dramatised as the ``worst theoretical prediction in the history of physics” \parencite[p.~187]{HobsonEfsthLasen}.\footnote{Initially propitious-seeming attempts to remedy the situation involved super-symmetry (which had an independent motivation, of course, besides its aesthetic appeal, see \textcite{Fischer2024-FISTPO-31}). No evidence for it is forthcoming. Experiments put extremely tight constraints on models. They render still viable supersymmetric extensions of the Standard Model of particle physics contrived---guilty of structural ad-hocness through fine-tuning of parameters.}
Following the promise of external coherence, we wind up in a cul-de-sac: lest we run into the aforementioned ``worst prediction in the history of physics'', we are forced to resort to increasingly implausible speculation and/or egregious other kinds of ad-hocness. The spectre of such liabilities curbs the pursuit-worthiness of the vacuum energy interpretation.\footnote{Recently, Unimodular Gravity has received some attention (e.g.\ \textcite{EarmanUNIMODULAR,EllisetalTRACEFREEE, EllisINFLATION}). It may be viewed as an especially conservative and simple approach in three key respects. First, in terms of its mathematical principles, it deviates from standard GR (\textit{sans} $\Lambda$) merely in postulating an absolute/non-dynamical volume element (e.g.\ \textcite{Stachel}), constant across all solutions. Secondly, the resulting field equations reproduce the \textit{full} Einstein field equation \textit{with} a cosmological constant---but as a model-dependent integration constant, rather than a constant of nature. And thirdly, it doesn't require any revisions of quantum field theoretical assumptions or calculations about the quantum vacuum: in particular, its standardly calculated value can be retained without leading to a conflict. By construction, Unimodular Gravity decouples the (quantum field theoretic) vacuum energy from gravity: the former no longer contributes to the total energy-momentum that generates general-relativistic gravity. Thereby, the theory ``resolves the problem of the discrepancy between the vacuum energy density and the observed value of the cosmological constant” (Ellis et al.\, 2010, p.1). 
Since the vacuum energy interpretation of $\Lambda$ rests on, as we stressed, contentious premises, the value of thus ``explaining away” its putative conflict with zero-point energy calculations seems questionable, however. Enthusiasm for unimodular gravity may be further curbed by the syntactic cost of its extra postulate, i.e. the postulate of an absolute volume element.
}
Fortunately---or, rather: further confounding an unambiguous appraisal of this approach's pursuit-worthiness---the interpretation of the cosmological constant in terms of vacuum energy contributions isn’t inevitable: as far as we know, we don't \textit{have to} adopt it. It rests on several controversial assumptions and has repeatedly and credibly been called into question (e.g.\ \textcite{Rugh2002-RUGTQV-2, Bianchi:2010uw, Koberinski_etal_2023, Schneider2020}).

The project of ``determining $\Lambda$" through quantum field theoretic considerations, if successful, would have forged a particularly strong link between GR and quantum field theory. But also a weaker link would be desirable, and cure some of $\Lambda$'s perceived ad-hocness (in the form of lack of coherence): the integration into a shared conceptual and methodical framework.

Regrettably, our hopes are dashed once more: GR with a cosmological constant fails to be ``natural". That is, GR with a cosmological constant can’t be treated as ``low-energy quantum gravity theory” \parencite{Wallace2022-WALQGA}, i.e.\, an effective quantum field theory---the powerful framework enabling the stellar predictive successes of QED or QCD. This may be lamentable (and diminish the attractivity of the cosmological constant approach, cf.\ \textcite[Ch.~IX.E]{Bartelborth1996-BARSE-2, Falkenburg2012-FALPUO}): it forecloses a more unified treatment of quantum field theories and GR (see e.g.\ \textcite{Williams2015-WILNTA-30}). On the other hand, \textcite{KoberinskiForthcoming-KOBAT} have persuasively argued that this failure of naturalness should be regarded as a feature, not a bug: the inapplicability of the Effective Field Theory framework isn’t anything inherently problematic, or antecedently warranted; the failure of naturalness, they urge, should be considered a reductio of that assumption. Again, in consequence, the attempt to ameliorate $\Lambda$'s external coherence fails: it defies integration into the larger framework of Effective Field Theories.

\subsubsection{Novelty}

After these disappointments with respect to external coherence, let's move on to more uplifting facets of $\Lambda$. In addition to its natural place within relativistic cosmology and simplicity, pursuit of the cosmological constant approach to DE is also commended by a weaker---primarily qualitative---variant of predictive or explanatory novelty: what we'll refer to as ``novelty", a theory’s capacity for novel ways to account for certain phenomena (see e.g.\ \textcite[Ch.~3]{Carrier1988-CARONF-2, Douglas2013-DOUSOT, Schindler2018-SCHTVI-5}). We’ll take a theory’s novelty to indicate greater promise than one lacking it (cf. \textcite[Sect.~7]{Wolf_Duerr_2023}). It provides a modicum of epistemic assurance that, in principle, the theory in question has the resources to account for phenomena of interest. Such tentative intimations of a theory's problem-solving power boost its pursuit-worthiness. They also increase the plausibility that the theory might, at some point, pass the usual evidential standards of acceptable theories (typically in the form of precise explanatory or predictive achievements). For our purposes, we'll rest content with a more qualitative sense of predictions and explanatory achievements, as seems apposite for not fully mature theories or research on the cusp of the unknown. 

Most germane for the case of the cosmological constant is what, with respect to predictions, has been dubbed “problem-novelty”. Problem-novel (qualitative) accounts ``don’t belong to the problem-situation which governed the construction of the hypothesis'' \parencite[p.~2]{Gardner1982-GARPNF}. That is, problem-novel features of a theory don’t belong to the class of problems that the theory’s inventor considered their theory responsible to solve. 

On this indicator of pursuit-worthiness, the cosmological constant approach to Dark Energy scores well (cf.\ for instance, \textcite{Earman2001-EARLTC, ORaifeartaigh:2017yby}). A non-vanishing value of $\Lambda$ counts as problem-novel in several regards:\footnote{Historically, Einstein’s original (1917) motivation for introducing the cosmological constant is curious: it provides likewise a problem-novel motivation for its non-vanishing value—albeit a specious one. In virtue of $\Lambda$, Einstein hoped to implement---or rather: rescue---his Machian intuitions (see e.g.\ \textcite{Smeenk2013}). Ironically, his hopes were soon dashed (ibid.), and he abandoned Mach’s principle, alongside $\Lambda$.}
\begin{itemize}
\item Until the 1960s many cosmological models \textit{sans} cosmological constant predicted an age of the universe younger than the oldest known stars. Before improved measurements of the Hubble constant in the late 50s  somewhat\footnote{Between the 1970s and 1990s, the so-called ``Hubble Wars" were waging, a dispute over two rivalling determinations of the Hubble constant. For the higher value of $\Lambda$, the Age Problem re-surfaced. The 1990s saw a resolution, as well as the emergence of ever stronger evidence for an accelerating universe---(at least) effectively described by a non-vanishing cosmological constant (see e.g.\ \textcite[p.~319]{Peebles2020}).} ``eased" (\textcite[p.~16]{10.1063/1.4902842}) this so-called Age Problem, a promising response was to allow for a positive $\Lambda$. This, in fact, was integral part of the Eddington-Lemaître model, the first Hot Big Bang model of the universe (e.g.\ \textcite[Ch.2]{Kragh1996}). It postulated a three-stage cosmic evolution: an initial phase of deceleration and a final phase of $\Lambda$-dominated accelerated expansion, with a ``loitering'' phase in the middle during which the gravity-induced deceleration is compensated for by $\Lambda$'s repulsive effects. This affords the time needed for the right age of the universe.   

\item Another appealing feature of the model, and of including a non-zero cosmological constant more generally (e.g.~\textcite{Carroll:1991mt}), ``was that it also offered a possible mechanism for the formation of galactic structures, a phenomenon that presented a formidable puzzle [...].” \parencite[p.~87]{ORaifeartaigh:2017yby}. The Eddington-Lemaîte model, thanks to its $\Lambda$-induced phase of stagnation, struck the requisite balance between gravitational forces and expansion needed for adequate structure formation.
\item A strong \textit{theoretical}\footnote{Note that the argument was articulated while inflation could still count as little more than a promising speculation, without compelling evidence (see e.g.\ \textcite{Smeenk:2017uof,Smeenk:2019gaining}). For \textcite[p.~3406]{Ratra:1987rm} this theoretical motivation is nonetheless strong for inflation ``offers the only known, reasonable, explanation for the remarkable homogeneity of the Universe within our horizon”.} motivation for $\Lambda \neq 0$ stems from the theory of cosmic inflation (see e.g.\ \textcite{Guth:2005zr}). That is, given the universe’s low observed mass density—including (dynamically inferred) Dark Matter---and inflation, which leads us to expect a flat universe, we must postulate additional effective energy contributions to make up the difference \parencite{Peebles_Ratra_1988}. Within GR, this is most naturally incorporated via the cosmological constant. 
\item Several variegated methods exist for determining the characteristic ingredients
of the $\Lambda$CDM-model (see e.g.\ \textcite{Smeenk2018}). Different methods, drawing on diverse phenomena and physical regularities, differing especially with respect to sources of systematic errors,
in particular allow for more or less independent ways to determine the cosmological constant. They probe different segments of the cosmic expansion history and/or aspects of cosmological growth structure (say, redshifts and distances accessible via measured data for supernovae, or via data from the CMB, see e.g.\ \textcite[Sect.~7]{Frieman:2008sn}). It’s therefore plausible to interpret the resulting determinations---the (ideally converging/consilient) value for $\Lambda$, as inferred from those outcomes---as emanating from distinct problem contexts.\footnote{Typically this consilience is considered in the context of acceptance, i.e.\ as a form of evidence, known as Perrin-style overdetermination (see e.g. \textcite{Smeenk:2017uof}). Consilience, if used in the context of \textit{pursuit}, is nothing but a form of problem-novelty warranting further study.} 

\item Perhaps the most astounding problem-novel achievement---plausibly even qualifying as a quantitative prediction---involves a famous anthropic argument for a positive cosmological constant \parencite{Weinberg:1987dv}: placing compatibility with existence of life as a constraint on viable cosmological models---surely a problem-context very different from what cosmological modelling usually deals with!---he obtained an estimate for $\Lambda$ within a few orders of magnitude of the value indicated by current data.  
\end{itemize}

\subsubsection{Fecundity}

Let’s close with the last major theory virtue of the cosmological constant approach to Dark Energy: fertility (or fecundity)---the ability to engender theoretical (and, ideally, empirically successful) innovation. \textcite[p.~7]{Smeenk:2023tgk} express representative reservations: ``(i)t is clear why treating $\Lambda$ as a true constant is unappealing. It is sterile, and fails to generate further consequences that can be pursued through theory or observational programs. It represents a dead end rather than a step towards further iterative refinements.” 

We tend to concur that pursuing or exploring $\Lambda$ may be somewhat sterile with respect to spawning \textit{fundamental} theoretical innovation. Yet, innovation needn't be limited to such fundamental revisions (see e.g.\ \textcite[Ch.~II-IV]{Kuhn1962-KUHTSO-3}). We'd like to push back against Smeenk and Weatherall's verdict at the declared level of generality: pursuing $\Lambda$ isn't sterile in terms of yielding important scientific insights (which in turn might prompt further theoretical developments).  

First, a non-vanishing cosmological constant has plenty of highly non-trivial implications well worth studying. The cosmological constant enters astrophysical and cosmological model-building, e.g.\ in galaxy formation or evolution processes, as a background assumption (see e.g.\ \textcite{Carroll:2000fy, Lopez-Corredoira:2017rqn}). Insofar as one judges such developments progressive and fertile---as we do---a share of the credit plausibly trickles down to $\Lambda$ as one of their premises.  

Furthermore, a non-vanishing cosmological constant---or an effective description mimicking its dynamical effects---also has dramatic, \textit{direct} consequences for the fundamental mathematical description of Black Holes or gravitational waves (see \textcite{Ashtekar:2014zfa, Ashtekar:2017dlf} for a review). For instance, nigh-universally the literature on gravitational waves assumes $\Lambda=0$. But ``(i)t turns out that the presence of a positive $\Lambda$ has a deep conceptual impact [...] because the limit $\Lambda \rightarrow 0$ is discontinuous'' \parencite[p.~1]{Ashtekar:2017dlf}. By the same token, standard Black Hole physics \textit{isn't} compatible with a non-vanishing cosmological cosmological constant---at least sufficiently far away from the Black Hole! Even if a non-vanishing $\Lambda$ turns out not to impinge upon astrophysical applications (usually presupposing $\Lambda=0$), such a result wouldn't diminish the importance and value of the demonstration that it doesn't.\footnote{It deserves to be underlined that also time-travel scenarios, such as the Gödel solution, occasionally rely on the cosmological constant (albeit usually with a negative sign). Insofar as one considers the exploration of such exotic scenarios non-trivial insights worth studying, one has a more speculative---though undoubtedly intellectually thrilling---reason to pursue the cosmological constant approach.}

The $\Lambda$CDM-model is nigh-universally hailed as the simplest way of accommodating Dark Energy phenomenology, both syntactically (i.e.\ as far as mathematical complexity is concerned) and in terms of ontological parsimony (i.e.\ as far as the number of (kinds of) entities is concerned that it postulates). These forms of simplicity pay off in further investigating the consequences of a non-vanishing $\Lambda$: they facilitate model-building and the testing of models. Those sympathetic to the spirit of Popperian falsificationism (such as e.g.\ \textcite{EllisSilk, EllisStandard}) will welcome \textit{this} kind of fertility for tests---perhaps a form of fertility for negative insights---as no less important than its usual form. 

If such tests resulted in a falsification, it would be particularly informative: because $\Lambda$CDM is the most conservative and simplest proposal, little wiggle-room for \textit{minor} modifications remains (cf.\ \textcite{SHAFIELOO2014171,Turner}). Once a falsification showed up in some domain, we’d likely have a smoking gun of the domain and circumstances in which GR requires revisions; the latter one may expect to be fairly drastic.

\subsubsection{Overall assessment}

In sum, on the basis of the theory virtues it instantiates, the pursuit-worthiness of the cosmological constant approach to Dark Energy receives a checkered score:

\begin{itemize}
\item \textbf{Conservativeness}: Not only does the cosmological constant not ``minimally mutilate” \parencite{QuineLogic} the existing paradigm of gravitational physics, GR; qua free parameter, $\Lambda$ forms part and parcel of it.
\item \textbf{Coherence}: $\Lambda$ scores high on \textit{internal} coherence as it naturally follows from the general-relativistic framework and is in excellent agreement with variegated, independent data.

As a free parameter, however, it’s an inherent source of ad-hocness; the latter is exacerbated by seeming fine-tuning, betraying what one may deem an explanatory foible. 

In terms of \textit{external} coherence, the connection with the rest of physics, quantum field theory in particular, remains unclear at best, violently inconsistent at worst.   
\item \textbf{Simplicity}: A non-vanishing $\Lambda$ is nigh-universally recognised as the simplest account of Dark Energy phenomenology on the main standard construals of simplicity.  
\item \textbf{Novelty}: A non-vanishing $\Lambda$ has independent motivations and rationales.
\item \textbf{Epistemic caution:} To minimise the uncertainties inherent in any physical modelling, and in light of the plurality of motivations, epistemic prudence counsels us to countenance a non-zero value---to explore this physical possibility. 
\item \textbf{Fertility}: The cosmological constant hasn’t generated much direct fundamental theoretical innovation. It has however several non-trivial implications, direct and indirect.  Within the $\Lambda$CDM-model, as a presumed \textit{background} theory, it's presupposed by more specific cosmological ideas. Exploring the consequences of $\Lambda \neq 0$ is likely to reveal far-reaching ramifications for standard astrophysics. 
\end{itemize}  
As one would expect of GR as a phenomenological framework (cf.\ \textcite{Lehmkuhl2019}; \textcite[Ch.1]{Kiefer}), the theory virtues that the cosmological constant approach most notably instantiates strike us as pragmatic (rather than uncontroversially epistemic-evidential): its conservativeness and simplicity enhance its testability. Pursuit of $\Lambda$ should (and can do little more than) focus on testing its consequences: tests---whatever their outcome---will be especially informative, both for probing GR’s domain of applicability, as well as for stimulating further theorising beyond orthodoxy.

\subsection{Modifying Gravity: $f(R)$ Gravity}

$f(R)$ Gravity epitomises the modified gravity approach to DE: it changes GR's gravitational dynamics (i.e.\ the dynamics for the metric field), whilst retaining the way gravity couples to ordinary (non-gravitational) matter.

\subsubsection{Simplicity and conservatism}$f(R)$ Gravity is usually introduced as the most conservative (or ``minimally mutiliated” in Quine’s parlance), or simplest extension of GR. We can parse this out in terms of mathematical and physical principles. 

First, \textit{mathematically}, $f(R)$ remains---like GR---within the framework of Riemannian geometry. By construction, $f(R)$ Gravity conforms also to GR’s further salient formal-mathematical properties, such as general covariance, absence of absolute objects, or reliance on the Riemann scalar for the action’s gravitational Lagrangian (rather than, say, the Riemann tensor, as countenanced proposals discussed in \textcite[Sect.~4.2]{Clifton_etal}). 

Secondly, $f(R)$ Gravity preserves GR’s key \textit{physical} assumptions: first, and foremost, the coupling of the gravitational variable---the spacetime metric---to the matter variables. The coupling scheme carries over verbatim from GR to $f(R)$ Gravity: universal, direct coupling to matter in the action’s matter part, and ``minimal coupling” between gravity and matter.
This ensures that the energy-momentum conservation law, another key principal of GR, continues to hold.
By the same token, two versions of the Equivalence Principle (see e.g.~\textcite{Lehmkuhl2021-LEHTEP}) are satisfied—the Weak one (asserting the universality of free-fall, and the fact that test-particles follow geodesics of the metric), and the Einstein Equivalence Principle (according to which, roughly, so long as a system’s self-gravity---tidal effects in particular---remains negligible, one can’t locally detect the presence of gravity if the system is in gravitational free-fall).\footnote{In contrast to GR, $f(R)$ Gravity \textit{violates} the Strong Equivalence Principle (the extension of the Einstein Equivalence Principle to systems with non-negligible self-gravity) through its presence of fifth forces. Note that $f(R)$ Gravity achieves consistency with experimental constraints on the validity of the Strong Equivalence Principle through adjustment of its free parameters. Unlike the Weak and Strong ones, this version of the Equivalence Principle is arguably best viewed as a \textit{theorem} rather than a (constitutive or heuristic) principle for GR.} 

As a result of retaining GR's principles, $f(R)$ Gravity inherits a key attraction from GR: rather than attributing it to additional entities, or types of entities (e.g. forces, mediated e.g. by scalars, as in quintessence models, see below), $f(R)$ Gravity conceptualises Dark Energy phenomenology as a manifestation of spacetime structure, in complete analogy to the way that GR conceptualises the perihelion advance or light deflection as manifestations of GR’s non-Minkowskian spacetime structure (cf.\ \textcite[Ch.~IX]{LehmkuhulTHesis}). Dispensing with the need for other entities, $f(R)$ Gravity thus retains GR’s qualitative parsimony.

Simplicity and conservativeness are typically viewed as a \textit{pragmatic} theory virtues: divorced from claims to truth or support, they are arguably invoked for reasons of convenience (or, occasionally, even tractability tout court). In terms of our cost/benefit rationale for pursuit-worthiness, they lower the costs of investigating theories instantiating them: ceteris paribus, simple and conservative theories require fewer cognitive resources for their exploration. Furthermore, insofar as it pertains to the salvaging of empirically well-corroborated principles---as in the case of the Equivalence Principle(s)---conservativeness also increases pursuit-worthiness by lowering inductive risk. 

Even as a pragmatic indicators of pursuit-worthiness, simplicity doesn’t as unambiguously distinguish $f(R)$ Gravity as one might initially think. $f(R)$ Gravity’s simplicity and conservativeness in the above sense are set off against losses of \textit{syntactic} simplicity (i.e.\ structural complexity). 
First, as a family of theories, corresponding to all possible choices of the function $f$, $f(R)$ Gravity contains infinitely many free parameters. Free parameters, however, are tantamount to forms of ad-hocness; the latter in turn is plausibly construed in terms of diminished coherence (see \textcite[Sect.~4][]{DuerrWolf_2023, Schindler2018-SCHTVI-5} for an elaboration)---an epistemic defect. Unless additional arguments pare down the form of $f(R)$, $f(R)$ Gravity’s conservativeness and simplicity collide with the demand for coherence, the shunning of ad-hocness. 

Secondly, although $f(R)$ Gravity, thanks to essentially the same mathematical and physical principles as GR, seems to be only modestly less simple when formulated at the level of an action principle, $f(R)$ Gravity’s \textit{field equations} are mathematically much more complex than those of GR. The latter---knotty enough!---are ``second order” (i.e.\ containing only second derivatives). Those of $f(R)$ Gravity, by contrast, are fourth order. Not only do significant mathematical complications ensue. Furthermore, we also need more input---initial value data---to extract quantitative predictions from $f(R)$ Gravity models. Arguably, this diminishes their explanatory/predictive power.     

Both losses of simplicity are entangled with another theory vice that $f(R)$ Gravity instantiates, specifically if $f(R)$ Gravity is intended as a solution to the Dark Energy problem. While $f(R)$ Gravity is capable of reproducing cosmic acceleration in principle,
to achieve these results, suitable functions $f(R)$ must be carefully \textit{engineered}. Such fine-tuning---a form of what \textcite{DuerrWolf_2023} label ``structural ad-hocness"---is unsatisfactory: by harnessing its infinitely many degrees of freedom (embodied in the ab initio free choice of $f$), it merely accommodates data; it doesn't predict  explain it in a forward-looking, novel way (cf.\ \textcite{Barnes2022}). $f(R)$ Gravity per se offers no guidance for further construction of a specific model (i.e.\ choice of $f(R)$). In this sense, $f(R)$ Gravity lacks what \textcite[p.~49]{Lakatos1978-LAKTMO} evocatively calls a `positive heuristic', ``(consisting) of a partially articulated set of suggestions or hints on how to change, develop the `refutable variants' of the research programme, how to modify, sophisticate'' $f(R)$ Gravity’s basic setup.

\subsubsection{Links to effective field theories?}

A second major motivation for $f(R)$ Gravity stems from the perspective of effective field theory. This denotes a general framework that on the one hand, allows the construction of low-energy limits, approximations of more fundamental quantum field theories. Even in the absence of such a fundamental theory, the framework permits the construction of an effective theory (observationally adequate at some scale). 

To extract finite quantities---and hence predictions---from such a theory, we need to ``renormalise” it---that is, remove its infinites. 
GR is famously non-renormalisable; however, introducing higher order geometric scalars can lead to renormalisable theories. Consequently, theories such as $f(R)$ gravity admit of more conventional quantisation \parencite[p.~2]{Sotiriou_Faraoni_2010}.

Note, however, that this argument only shows that renormalisability requires higher-order contributions to GR. $f(R)$ Gravity is one \textit{possible} higher-order generalisations—amongst other options. The effective field theoretical motivation is thus merely a weak reason for pursuit, based on coherence with the---highly successful---framework of effective field theories: the latter makes us expect curvature-containing “correction terms” to GR \textit{like}, but not necessarily limited to, those stipulated by $f(R)$ Gravity.

One may further question the effective field theoretical motivation on more general grounds that are specific to the Dark Energy problem---a problem that arises, of course, at cosmic scales. \textcite[p.~471]{KoberinskiForthcoming-KOBAT} have, to our minds convincingly, argued that ``(o)ne should reject the application of EFT [Effective Field Theory] methods to the far [Infrared] of cosmological spacetimes. Standard EFT methods depend on having specific types of background spacetime structure"---structure not present for the (effective) spacetimes associated with Dark Energy phenomenology.  

\subsubsection{Epistemic spin-offs}

We therefore take the third type of motivation for $f(R)$ Gravity to be the most compelling one: the spin-offs of studying $f(R)$ Gravity. \textcite[p.~4]{Sotiriou_Faraoni_2010} sum up the idea by underscoring $f(R)$ Gravity’s value as a toy-model, ``an easy-to-handle deviation from Einstein’s theory mostly to be used in order to understand the principles and limitations of modified gravity”. This third argument justifies the pursuit-worthiness of $f(R)$ Gravity through the insights that studying it is likely to yield---even if $f(R)$ Gravity ultimately isn’t judged adequate for describing our world. Pursuing $f(R)$ Gravity \textit{as a toy-model}, in other words, is justified by the collateral pay-off: it allows us to learn something about modifications of GR more generally. Thereby the pursuit of $f(R)$ Gravity enhances our understanding also of GR: it deepens our grasp of viability constraints of gravitational theories, the robustness of certain results (e.g.\ No Hair Theorems), or alerts us to phenomena that might serve as smoking guns of deviations from GR. 

$f(R)$ Gravity displays, already in simple models, many of the qualitative features and novel phenomenology, characteristic of higher-order generalisations---and in fact broader extensions---of GR. Phenomenologically, $f(R)$ Gravity exhibits, for instance, (vis-à-vis GR) extra excitation modes: gravitational radiation not only has a transversal tensor component (familiar from GR), but also longitudinally propagating scalar one, travelling at a speed below that of light. In the same vein, $f(R)$ Gravity implies ``fifth forces”, i.e. Yukawa-like correction terms to the Newtonian gravitational potential. But also on the more formal side, the utility of a---qualitatively rich---contrast theory to GR is obvious: via actual competitor theories, one can refine, and make stricter, the systematic testing of GR---for instance, what potential deviations from the standard theory’s predictions to look for. This is precisely what happened: studying $f(R)$ Gravity led to a deeper understanding of the intricacies and pitfalls of the PPN formalism, a framework for testing (and quantifying) deviations from GR (see \textcite[Sect.~2.5]{Sotiriou:2008ve}). 

\subsubsection{Overall assessment}
The pursuit-worthiness of $f(R)$ Gravity receives a mixed score, as far as its intrinsic virtues are concerned. The  pursuit is primarily justified extrinsically---by likely epistemic \textit{by-products}: 
\begin{itemize}
    \item \textbf{Simplicity and conservativeness:} These don't, in any obvious way, favour $f(R)$ Gravity over other Dark Energy proposals.
    \item \textbf{Coherence with Effective Field Theories:} It turns out to be weak; the adequacy of this perspective seems doubtful. We found $f(R)$ Gravity's heuristic power for accounting for Dark Energy phenomenology severely limited.
    \item \textbf{Epistemic spin-offs}:  If, however, one grounds the pursuit-worthiness of $f(R)$ Gravity \textit{extrinsically}---in the pay-offs of studying it, without treating it as an adequate description of reality---$f(R)$ Gravity scores highly: as a toy-model, it enhances our understanding of modifications of GR, and of GR itself.

\end{itemize}

\subsection{Quintessence}
Quintessence opts for keeping GR (\textit{without} a cosmological constant---or rather one postulated to vanish) intact, but revises a background assumption: the assumption that \textit{only} ``ordinary matter” (including Dark Matter) sources gravitational dynamics. Alongside normal matter, quintessence models posit a new type of matter, the ``quintessence field''.

\subsubsection{Ease of model-building}

The quintessence ansatz is  attractive for reasons of explorative theory/model construction. Thanks to ample ``experience with the tools commonly used to develop theories beyond the standard model of particle physics or Einstein’s theory of gravitation'' \parencite[p.~215]{Caldwell_2000}, it gives cosmologists something to work on, delineating a straightforward line of inquiry, under good cognitive control, sufficiently well-constrained, and at the same time sufficiently versatile. 
Quintessence accrues such ease in tractability from two aspects. 
First, a scalar is arguably the mathematically simplest type of matter field one can consider in general-relativistic field theory. Almost all GR textbooks treat it as a pedagogical example. Applied to cosmology, its incorporation leaves the basic form of the dynamics unchanged. The Friedmann equations are preserved. The quintessence field’s equation of motion itself is likewise familiar: the Klein-Gordon equation with a potential. Also the evolution of perturbations—the dynamics governing the propagation of small fluctuations in the quintessence field—remains mathematically manageable, well within the domain of conceptually standard mathematical techniques.
Secondly, the cosmological effects of scalars in cosmology have been studied extensively. Some of GR's oldest extensions, so-called scalar-tensor theories \parencite{Faraoni:2004pi, Fujii:2003pa}, involve scalars.
Quintessence models explicitly strive to mimic another successful application in cosmology \parencite[p.~215]{Caldwell_2000}: the inflaton, i.e.\, the scalar field driving cosmic inflation in the very early universe. 

The intimate familiarity with a minimally coupled scalar enables researchers to tap existing cognitive resources in pursuing quintessence models (cf.\ \textcite{Nyrup2020-NYROWD}). In the economic terminology of our Peircean account of pursuit-worthiness, the costs for investigating quintessence models are low. What, then, does quintessence allow us to purchase? Three merits stand out: the explanatory power to address a fine-tuning issue, the prospect of a unified account of early \textit{and} late-time accelerated expansion of the universe, and certain empirical predictions. 
\subsubsection{Explanatory and predictive achievements}

The main allure of quintessence approaches lies in their explanatory power \parencite[p.~35]{Carroll:2000fy}: first and foremost, it offers a prima facie elegant solution to the ``coincidence problem”, a perceived fine-tuning of initial conditions required to account for current observations \parencite{Zlatev:1998tr}. 

At the heart of the coincidence problem lies the observation that the current energy densities of Dark Energy and matter are of the same order of magnitude. Since the two densities decrease at different rates over cosmic history, this is puzzling: the radiation energy density scales as $\rho_r \propto a^{-4}$, the matter energy density scales as $\rho_m \propto a^{-3}$, and the energy density one can associate with $\Lambda$ remains constant, $\rho_{\Lambda} \propto a^0$. At the beginning of a Hot Big Bang state, radiation would naturally dominate. But this energy density quickly becomes negligible as it drops significantly faster than matter. Throughout the lion's share of its history, the universe was dominated by matter. Only quite recently did the Dark Energy density eclipse the matter energy density, leading to the Dark Energy driven expansion that we find ourselves in now. Extrapolating back in time, ``it appears that their ratio\footnote{\textcite[p.~218]{Caldwell_2000} gives the estimated ``ratio of energy densities to 1 part in $\approx 10^{110}$ at the end of inflation”.} must be set to a specific, infinitesimal value in the very early Universe in order for the two densities to nearly coincide today, some 15 billion years later” \parencite[p.896]{Zlatev:1998tr}.
If instead\footnote{NB: \textcite[p.~115]{Amendola_Tsujikawa_2015} emphasise: ``(t)he coincidence problem is not specific to the cosmological constant. Almost all acceptable dark energy models [...] behave similarly to the cosmological constant.''} of a cosmological constant, we attribute the universe’s accelerated expansion to the effect of quintessence, we can circumvent the coincidence problem: so-called tracker potentials have been found where ``a very wide range of initial conditions rapidly converge to a common, cosmic evolutionary track'' \parencite[p.~59]{Steinhardt:1999nw}. 
Qualitatively,\footnote{For details see \textcite{Steinhardt:1999nw, Martin_2008, Caldwell_2000}.} quintessence models with tracking behaviour exhibit two features:
\begin{enumerate}
    \item The equation of state for the scalar field, $w_{\varphi}$, imitates, or ``tracks”, the behaviour of the energy component dominant during each cosmic epoch. When the universe is radiation-dominated ($w=1/3$),
    $w_{\varphi}$ will be less than or equal to 1/3. When the universe is matter-dominated ($w=0)$, $w_{\varphi}$ will be less than or equal to 0. When the scalar field becomes the dominant component, its equation of state approximates a cosmological $w \approx -1$: the universe enters a period of accelerated expansion.
    \item The tracking quintessence dilutes less rapidly than radiation or matter, so eventually it will dominate the universe's energy budget. But rather than remaining constant---as in the cosmological constant approach---$\rho_{\varphi}$ mimics the scaling of radiation and matter far more closely throughout most of the universe's history. It’s thus far less sensitive to these initial conditions. 
\end{enumerate}
As a result, the initial value for $\rho_{\varphi}$ needn’t be set as an extraordinarily specific input. Rather, it can take on a significantly larger range of values without disrupting the capacity of our cosmological models to accurately describe the evolutionary history of our universe. In quintessence models with this kind of tracking behaviour, initial $\rho_{\varphi}$ may vary by around 100 orders of magnitude \parencite{Steinhardt:1999nw, Martin_2008}. They thereby alleviate the fine-tuning and are robust in the sense that ``that a wide range of initial conditions is funneled into the same final condition'' \parencite[p.896]{Zlatev:1998tr}. This mirrors an explanatory benefit of inflation in the early universe \parencite[p.~10]{Steinhardt:1999nw}.\footnote{Insofar as advocates of quintessence rely on the \textit{analogy} with inflation, we’d like to pour a little cold water on it.  The (inflation-less) Hot Big Bang model’s fine-tuning problems require a judicious choice of initial conditions:
uniformity in the distribution of matter, a spatially flat geometry, and a-causal correlations in large-scale structure (see e.g.~\textcite{McCoy2015-MCCDIS}).  
One may baulk at the analogy due to a salient difference: the fine-tuning required for the empirical adequacy of the inflation-less Hot Big Bang model is at least qualitatively worse than the cosmic coincidence problem. The former is constituted by a \textit{triple} fine-tuning—for flatness, uniformity, and a-causal correlations. 
Quintessence, by contrast, is a solution to a \textit{single} fine-tuning problem (the cosmic coincidence problem).
Furthermore, what makes the postulate of a scalar in inflationary cosmology so attractive is a cornucopia of benefits (including successful novel predictions) that go above and beyond resolution of these fine-tuning issues \parencite{Wolf_Duerr_2023}. Quintessence models cannot boast of similar merits.}

Not only do quintessence models thereby demonstrate the capacity for explaining apparent fine-tuning; else, the latter would have to be swallowed as an enigmatic brute fact. Thanks to its robustness, the account that quintessence provides has a particular explanatory quality: insensitivity to, or invariance under counterfactual variation in, initial conditions renders the explanation \textit{``deep”} \parencite{Hitchcock2003-HITEGP, Ylikoski2010-YLIDEP}. Its main advantage is a pragmatic one for model-building: being relatively independent of contingent input that is typically epistemically inaccessible or at least vexed with uncertainty, such an explanation hedges the researcher’s epistemic bets. In terms of our economic analysis of pursuit-worthiness, quintessence thus offers a high epistemic gain (viz.\ an explanation of an otherwise puzzling anomaly) the cosmic coincidence problem---at a low epistemic risk (viz.\ by underwriting an explanation that is deep in the foregoing sense).

This ``deep'' explanatory resolution of the cosmic coincidence problem is frequently touted as a major achievement of quintessence models. But one may resist the enthusiasm: even if one concedes that its ability to explain a coincidence makes a theory pursuit-worthy (see e.g.\ \textcite[p.~4]{Carroll:2014uoa}; \textcite[Sect.~4]{Wolf_Duerr_2023}; \textcite[Sect.~3]{Wolf_Karim_2023}), it’s possible to contest that a coincidence \textit{problem} exists. The impression that the current ratio of Dark Energy and Dark Matter densities singles out a uniquely special time in cosmic history---today!---hinges on using \textit{redshift} as the time variable (used because it’s a directly observable quantity): ``(i)n the cosmic-time parametrization the densities of dark energy and dark matter are of a similar order over a substantial fraction of cosmic history, not just 'recently'" \parencite[p.~4]{Velten:2014nra}. From this vantage point, the ``why now?”-question, underlying the cosmic coincidence problem, seems much less pressing. On the other hand, the present value of the Dark Energy density remains highly constrained by both structure formation and the age of universe. The dynamical mechanisms characteristic of quintessence models still allow for a significant relaxation of this value on the choice of initial conditions \parencite[Sect.~IV]{Velten:2014nra}.

A second alleged explanatory benefit of the quintessence approach resides in its capacity for a unified account of early-time and late-time accelerated expansion. It’s possible to construct quintessence models where the scalar is responsible for both inflation and Dark Energy phenomenology (e.g.\ \textcite{Peebles:1998qn}).\footnote{Early models of quintessence by \textcite{Peebles_Ratra_1988} explicitly commenced from the idea of extending the inflaton’s effects to late-time expansion.} Such a unified account would indeed be desirable. Yet, \textit{in their present forms}, their epistemic gains are elusive: the unification seems merely formal; a potential has to be meticulously concocted to do the work. The key defect is patent: physically motivating a potential that reproduces inflationary and late-time expansion is even knottier than motivating a potential for either individually. We’ll zoom in on that challenge---a physical justification for the scalar’s dynamics (or potential)—further below.\footnote{Similar reservations apply to attempts at a unified account of Dark Energy and Dark Matter.}

A third achievement of quintessence models concerns a novel, testable prediction: the Dark Energy equation of state will deviate from $-1$. Future measurements will further probe the effective equation of state associated with Dark Energy phenomenology. Deviations from $-1$ will indicate the break-down of the cosmological constant approach. \textcite[p.~228]{Caldwell_2000} judges ``(t)he prospects for decisively testing the quintessence hypothesis [to be] excellent”.

However, the empirical situation is made more complicated by subtle degeneracy issues, i.e.\ evidential underdetermination (see e.g.\ \textcite[Sect.~4]{Steinhardt:2003st}). From the perspective of this paper’s philosophical theme---theory virtues as indicators of pursuit-worthiness---a more fundamental methodological defect stares one in the face: so long as no compelling reasons are forthcoming for narrowing down the choice of potentials for the scalar, the testability of quintessence is curtailed. As \textcite{Wolf_Ferreira_2023} argue, even if we were to detect unassailable evidence of a temporal variation in the Dark Energy equation of state, such evidence would do almost nothing to constrain the form of the quintessence potential: numerous distinct quintessence models can arbitrarily saturate observable parameter space for $w(a)$.\footnote{This underdetermination at the level of the equation of state's phenomenological behaviour isn't restricted to quintessence models. It also affects modified gravity proposals as those will share much of the same phenomenological parameter space. One may hope to break the degeneracy between quintessence and modified gravity proposals through checks of non-minimal coupling (see e.g. \textcite{Burrage:2017qrf,Wolf:2019hun}).} We concur with \textcite[p.~47]{Ellis:2006fy}: even if deviations from the cosmological constant prediction were observed, quintessence models shouldn't be favoured ``until we have either in some independent experimental test demonstrated that matter of this form exists, or have theoretically shown why this matter or field has the form it does in some more fundamental terms than simply a phenomenological fit”.

On the other hand, quintessence models, with their predicted deviation from the cosmological constant approach, strongly motivate tests of the $\Lambda$CDM model. Moreover, quintessence models sharpen attention to where to look for a conceivable break-down of the standard model and how to assess such a break-down if it did appear.
To this end, the latest results from \textcite{DESI:2024mwx} hint at some dynamical behaviour in the Dark Energy equation of state. Should further investigations continue to point in this direction or  provide more conclusive evidence for such temporal variation, in contrast to the predictions of the standard model with $\Lambda$, quintessence provides a natural framework within which to explore these results and investigate the nature of Dark Energy exhibiting such behaviour.   

In short, we regard the predictions that quintessence models make for deviations from the standard model as an indirect benefit: they enhance the latter’s testability. This befits an approach intended as a class of phenomenological models, rather than fundamental theories (e.g.\ \textcite[p.~3408]{Peebles_Ratra_1988}).

\subsubsection{Lack of coherence}
Having reviewed the main virtues, let’s now scrutinise the key drawbacks of quintessence models. \textcite[p.~9]{DurrerMaartens} circumscribe them: quintessence models ``require a strong fine-tuning of the parameters of the Lagrangian to secure recent dominance of the field [...]. More generally, the quintessence potential, somewhat like the inflaton potential, remains arbitrary, until and unless fundamental physics selects a potential. There is currently no natural choice of potential.”

The demur can be couched in terms of our adopted indicators of pursuit-worthiness, theory virtues/vices: quintessence models are marred by deficient coherence. 
Quintessence models score poorly on \textit{internal} coherence:

\begin{itemize}
\item All quintessence models involve additional free parameters and free functions.
Theory-\textit{inherent} resources don’t furnish reasons for their specification; they are stipulated, by fiat, to display the desired features or satisfy empirical constraints. Hence, the free parameters and functions constitute serious lack of coherence; they spawn ad-hocness (cf.\ \textcite{Schindler_2014}; \textcite[Ch.~5]{Schindler2018-SCHTVI-5}). 
\item Also the formal-structural characteristics of the \textit{dynamics} of quintessence models may raise eyebrows. Apart from queries concerning the justification of the specific potential for quintessence models, the fact that the quintessence field only couples to gravity, but not to other matter fields, is unusual: all other forms of known matter can, at least in principle, interact with each other. One may legitimately hanker after a deeper reason for it; none is given, though. Instead, it's stipulated \textit{solely} to avoid conflict with (so-called fifth force) experiments.

\end{itemize}
On \textit{external} coherence---their connection with other areas of physics---quintessence proposals fare little better:
\begin{itemize}
\item As far as the \textit{existence} of scalars is concerned, quintessence models cohere with background knowledge in particle physics: at least one scalar, the Higgs boson, is known to exist. Furthermore, it’s widely accepted in high-energy cosmology that the inflaton, typically presumed to be a scalar field, is another candidate---even though it’s empirically not (yet) verified.

More generally, as \textcite[p.~2502]{Steinhardt:2003st} (our emphasis) stresses, quintessence models are ``constructed from \textit{building blocks} that are common to most quantum field theories". \textit{Elements} of quintessence models thus likewise cohere with background knowledge. 

\item Nonetheless, one shouldn’t exaggerate that coherence. Typical potentials for quintessence are motivated by \textit{similarity} to those from other areas. But the connection remains far from tight. 

A major source of \textit{tension} in fact stems from the mass associated with the quintessence field. Here, it's instructive to first recall the hierarchy problem in particle physics (construable as a coherence issue, see \textcite{Fischer2024-FISTPO-31, Fischer2023}): why is the Higgs boson's mass so much lighter than expected, given the quantum corrections it should receive from high-energy scales? The measured value requires an unnatural fine-tuning of parameters in the Standard Model of particle physics. 
The energy scale of Dark Energy implies that the mass scale of the quintessence field is $m_\phi \leq H_0 \sim 10^{-33} \mathrm{eV}$. This is \textit{incredibly} light! It’s obscure how such a nearly massless scalar field would be protected from radiative corrections at higher energies (including from all the established standard model fields). The standard hierarchy problem ``amounts to asking why the Higgs mass [now known to be $\sim 10^{11} \mathrm{eV}$] [...] should be so much smaller than the grand unification/Planck scale, $10^{25}-10^{27} \mathrm{eV}$" \parencite[p.~31]{Trodden:2004st}. Comprehending how a scalar field could have a mass $m_\phi \sim 10^{-33} \mathrm{eV}$ is significantly more challenging. Consequently, quintessence generates grave incoherence with respect to our background knowledge regarding quantum field theory and the Standard Model of particle physics.

\item As far as coherence with other, more speculative theories is concerned, it's worth noting that on the one hand, string theory---arguably the strongest candidate for a quantum theory of gravity (see e.g.\ \textcite{Dawid:2013maa})---seems to face some difficulties in reproducing GR with a cosmological constant. This suggests that Dark Energy may be dynamical. On the other hand, it would be hasty to embrace this as a victory for quintessence. To the contrary: ``[...] string theory does not naturally lead to scalar fields with the energy scale required to be a candidate for quintessence'' \parencite[p.~4]{Heisenberg:2018yae}.

\end{itemize}

\subsubsection{Overall assessment}
Appraising the pursuit-worthiness of quintessence approaches to Dark Energy, we end up with the following evaluation:
\begin{itemize}
\item \textbf{Familiarity and simplicity:} A simple and familiar ansatz, whose cosmological ramifications are straightforward to handle, quintessence models are easy to construct and analyse. The cognitive costs of exploring quintessence are relatively low. 
\item \textbf{Explanatory power:} Quintessence models promise a solution to the cosmic coincidence problem, prima facie an ``ultrafine-tuning problem for initial conditions” \parencite[p.~225]{Caldwell_2000}. Quintessence explains the coincidence in an epistemically gratifying manner. However, whether the coincidence problem is a fine-tuning problem that screams for an explanation seems debatable.

Through a suitable choice for the quintessence potential, quintessence provides, in principle, a unified account of both late-time---Dark Energy-related—and early-time---inflation-related---accelerated cosmic expansion. The inferential links in such a unified approach, however, aren’t tight; the unification seems artificial. The epistemic gain it yields is limited.  
\item \textbf{Testability:} Quintessence models make predictions, distinct from those of the cosmological constant approach to DE. Albeit at present eluding detection, it’s reasonable to assume that in the not-too-distant future tests of specific models will be possible; whatever the outcome they will yield valuable insights. 
\item \textbf{Coherence \& non-ad-hocness:} On the one hand, at least one scalar is known to exist: the Higgs boson. With the inflaton, one has yet another not implausible—but by no means corroborated---candidate. Hence, as a matter of principle, postulating the quintessence field as a scalar coheres with our wider web of beliefs in standard particle physics and more speculative theories.

On the other hand, quintessence models are compromised by notable ad-hocness---worse than the cosmological constant approach. They rest on physical principles, and introduce new degrees of freedom and finely-tuned parameters, for which no convincing independent justification seems forthcoming. This vice arguably diminishes the appeal of pursuing quintessence.
\end{itemize}
In conclusion, we concur with \textcite[p.~9]{Durrer:2008in}: ``(q)uintessence models do not seem more natural, better motivated or less contrived than $\Lambda$CDM. Nevertheless, they are a viable possibility and computations are straightforward. Therefore, they remain an interesting target for observations to shoot at.” That is, investigating quintessence models primarily serves \textit{explorative} purposes: in line with the attitude expressed by many main proponents of quintessence (e.g.~\textcite[p.~3408]{Ratra:1987rm}), it allows researchers to reconnoitre the phenomenological territory, and develop and refine effective descriptions. The main epistemic gain one may expect from pursuing quintessence models is indirect: they enhance the testability of the current cosmological standard model---with its cosmological constant $\Lambda\neq 0$---and facilitate the search for empirical deviations from the latter, all at minimal cognitive costs.

\subsection{Inhomogeneous models} 
They relax some of the auxiliary/background assumptions underlying standard cosmology, viz.\ those concerning the distribution of matter.

\subsubsection{Internal coherence and conservativeness}

Inhomogeneous models vie with the cosmological constant approach for being the most conservative Dark Energy account. They retain GR in its standard form, i.e.\ Eq.~(\ref{EFE}), without citing exotic forms of energy. Models of this type seek to explain Dark Energy phenomena exclusively through the resources of GR and conventional theories of (dark and baryonic) matter. 
Consequently, both backreaction models and void models are as ontologically parsimonious as the cosmological constant. This conservativeness and parsimony---explicitly extolled in the physics literature (e.g.\ \textcite[p.~76]{Buchert_2011}; \textcite[p.~2]{Celerier:1999hp}; \textcite[p.~810]{2002PThPh.108..809I})---can be cashed out in terms of the resulting internal coherence of these proposals. Further below (\textbf{§3.4.3}), we'll unpack severe qualifications of such positive first impressions.

\subsubsection{Epistemic spin-offs}
Pursuing such theories also has important epistemic spin-offs. First, we \textit{know} both that the assumption of homogeneity necessarily breaks down at some scale in order to account for cosmic structure, and that GR is inherently non-linear. Consequently, it's certainly worth investigating what effects such considerations might have on cosmological dynamics. Physics is replete with---often astonishing!---phenomena due to non-linearities (e.g.\ turbulance in fluid dynamics, or solitons in particle physics or nonlinear optics). Non-linearities are also crucial for gravitational waves astrophysics \parencite{LIGOScientific:2016lio, Will:2014kxa}. In short, non-linearities and inhomogeneities incontestably exist in the universe. Hence it's imperative to pin down \textit{the extent} to which these effects can be safely neglected in cosmological modelling.\footnote{See \textcite{Sarkar} some recent investigations into this question.} With \textit{that} objective in mind---an objective distinct from, and less ambitious than, finding a solution to the Dark Energy problem!---the line of inquiry that inhomogeneous cosmologies represent should be pursued.

A second kind of spin-off from exploring inhomogeneous models boosts their pursuit-worthiness more directly: they allow us to test the Cosmological Principle (asserting the homogeneous and isotropic distribution of matter on the universe's large scales). The latter is a foundational ab initio assumption of standard cosmology. Justifying it proves to be highly non-trivial and fraught with controversy (\textcite{BeisbartJung},\ \textcite{Beisbart2009}; \textcite[Sect. 4.2]{Ellis:2006fy}). It's all the more important to devise and perform empirical tests of the Cosmological Principle. The pursuit of inhomogeneous cosmologies, with their overt violation of the Cosmological Principle, spurs on and inspires such tests \parencite{Sarkar:2007cx, Ellis}. Furthermore, there have been persistent curiosities in the data that add further motivation to develop and explore cosmological models that relax the Cosmological Principle. See \textcite{Aluri:2022hzs, Secrest:2020has} for some recent work highlighting these features.

\subsubsection{Syntactic nightmare and external incoherence}

Both the backreaction and the void model variants of inhomogeneous cosmologies at first blush excel in terms of parsimony  and internal coherence. Closer inspection, however, discloses that not only do they introduce complexities that make assessing these proposals taxing; in some cases it's this syntactic complexity that elicits worries about substantial incoherence, both internal and external.

It's intuitive that an approach that critically relies on non-linearities will be significantly more complicated than one that doesn't. However, backreaction models also have some unique problems. They render assessing the programme particularly difficult. The most pressing issue concerns ambiguity in how one ``averages'' the inhomogeneities, corresponding to the arbitrariness of decomposing a general-relativistic spacetime into time and space. Different choices can lead to different outcomes regarding whether or not backreaction effects are significant enough to affect cosmological dynamics \parencite{Ishibashi:2005sj}. Another concern is rooted in poor epistemic control over the right \textit{amount} of correction terms: \textit{if} second order terms are relevant, why shouldn't even higher order terms be so as well (in contrast to linear perturbation theory, where all second and higher order terms are small and can be neglected)? The matter is further compounded by the fact that different orders plausibly induce non-trivial cancellation effects amongst each other (see e.g.~\textcite[p.~294]{Amendola_Tsujikawa_2015}; \textcite{Hirata:2005ei}). No consensus exists on a principled way of determining the order to which one must carry the corrections. Absent such a scheme, it remains somewhat opaque how such models can deliver informative predictions or even how one might objectively assess their viability.

In addition to such flaws of internal incoherence, one may fret also over potential \textit{external} incoherence: the successful application of standard cosmology on \textit{all} scales (except in the immediate vicinity of neutron stars or Black Holes) suggests the approximate validity of the linearly perturbed FLRW description.\footnote{A bit more in detail, the form of the perturbed FLRW metric indicates that large scales are described by the FLRW metric, while small scales are described by Newtonian gravity. Even on small scales with large matter overdensities where non-linear gravitational collapse occurs, the linearly perturbed FLRW spacetime metric remains a good approximation because the corresponding correction to the unperturbed metric remains small even here.} The fact that this approximation works so well would be mysterious if backreaction effects were the ultimate cause of cosmic acceleration \parencite{Ishibashi:2005sj}. Prima facie, one doesn’t expect $\Lambda$-imitating effects from GR’s non-linearities. Backreaction models thus don’t seem to naturally cohere with other astrophysical and cosmological knowledge.\footnote{cf. \textcite[p.17]{Buchert_2011} for a riposte to the claim of incompatibility, calling for efforts to develop generalisations of existing standard perturbation theory to conclusively settle the matter.}   

Void models---as inhomogeneous solutions to the Einstein field equations (\ref{EFE})---create problems of their own that undercut the models' initial promise. The form of the LTB metric employed in void models contains several free functions, additional variable dependencies, and new parameters. The model-building process thereby dramatically forfeits coherence: owing to this freedom, void models display substantive ad-hocness (in the sense explicated earlier). In many regards, the results from \textcite{Bull:2011wi} are representative. On the one hand, void models can provide a good fit for supernovae data. On the other hand, it's nearly impossible for void models to evade conflict with other cosmological data points, CMB radiation in particular.\footnote{More specifically, we would naturally expect a universe with large inhomogeneities to display large kinematic Sunyaev-Zel’dovich effects, whereby photons scatter off high-energy electrons found in galaxy clusters. This produces a Doppler shift in CMB photons that should create large anisotropies in the CMB.}
This effectively rules out ``simple'' LTB void models and would force one to resort to meticulously engineered and contrived specifications of free functions and parameters, leading to further structural fine-tuning and ad-hocness.  

Some further caveats demonstrate the inherent difficulty of producing, much less assessing, a potentially viable void model. For instance, most approaches assume that Earth is located at the \textit{exact} centre of the void. Relaxing this assumption, as any realistic model inevitably must, complicates interpretations of anisotropies in the CMB (see e.g.\ \textcite[p~.13]{Bull:2011wi} or \textcite{Alnes:2006pf}). Furthermore, perturbation theory---the effective computational means for quantitatively evaluating models---becomes ``significantly more complicated” (\textcite[p.7]{Bull:2011wi}, see \textcite{Clarkson:2009sc} for details) than in standard cosmology.    
Finally, for a \textit{late-time} void model to be viable---which, as an LTB model, only describes a matter/dust-dominated universe---we need to glue it to an earlier one to describe the universe’s \textit{radiation}-dominated phase \parencite[p.13]{Bull:2011wi}. Although a consistent matching is (in principle) possible, the complications and artificiality are palpable. Essentially, we lack good epistemic control over viable void models; moreover the cognitive costs of developing such models are unclear. 

The blemishes of void models aren’t confined to such an explosion in mathematical complexity and ad-hocness. Like backreaction models, they don’t seamlessly cohere with cosmological background knowledge: if the universe harboured large inhomogeneities, we’d expect distinct observational signatures, such as significant anisotropies in the CMB. However, we don’t see such signatures. Hence the models’ viability necessitates even more complexity and fine-tuning.  Furthermore, ``there is no valid mechanism at present to explain the formation of such huge inhomogeneities'' \parencite[p.~292]{Amendola_Tsujikawa_2015}. This is especially difficult to make sense of if the (currently favoured) inflationary mechanism for structure formation is correct. Inflation generically predicts that resulting density perturbations should be highly uniform in terms of their direction, location, and scale \parencite{Guth:2005zr}. Of course, one can introduce specific tuning into inflation to engineer ``under-densities” that evolve into voids \parencite{Afshordi:2010wn}. But this would, again, come at the cost of injecting further arbitrariness into a proposal already saturated with free functions, fine-tuning, and ad-hocness.

\subsubsection{Overall assessment}
The pursuit-worthiness of inhomogeneous models lies primarily in their epistemic spin-offs. Intended as proposals \textit{for Dark Energy}, their appeal is hampered by their cognitive costs and queries about their viability. 
\begin{itemize}
\item \textbf{Internal coherence and conservativeness:} These models are conservative and coherent in that they utilise only notions inherent to GR, without invoking exotic entities. 
\item \textbf{Epistemic spin-offs:} Investigating inhomogeneous cosmologies stimulates the search for potential observational signatures of non-linearities, and more austere tests of the Cosmological Principle.
\item \textbf{Syntactic complexity:} These models introduce significant mathematical complexities. These cognitive costs even impede our ability to uncontroversially extract predictions or even assess the models' viability. 
\item \textbf{Lack of external coherence:} Neither backreaction nor void models sit comfortably with our expectations and cosmological background knowledge.  
\end{itemize}

\section{Conclusion}

\noindent Having been persuaded by our diagnosis of the Dark Energy problem as chiefly a crisis of underdetermined pursuit-worthiness, one might finally ponder: what methodological conclusions to draw? What advice for the way forward might philosophers of science purvey?

Our analysis suggests a complementary double strategy, primarily rooted in pragmatic considerations---vindicating already existing practice.\footnote{E.g.\ ``In the next 15 years, LSS and CMB surveys […] will have the potential to constrain dynamical dark energy and departures from GR at the few-percent level. This will either rule out a large swath of the interesting parameter space of dark energy and modified gravity models or yield another breakthrough in cosmology with the detection of departures from $\Lambda$CDM" \parencite[p.~118]{2016ARNPS..66...95J}.} One component aims at promoting theory pluralism, reminiscent of Feyerabendian anarchism (cf.~\textcite{Lloyd}; \textcite{Shaw:2017}; \textcite{Bschir:2015}; \textcite{Brad2021}): 
we should encourage the exploration of alternatives---also and especially beyond the mainstream. Thereby, through the existence of, and confrontation with, rivalling theories, we plausibly augment the testability of the current prevailing standard model of cosmology. The second component, reminiscent of the spirit of Wheeler's ``daring conservatism" \parencite{Furlan:2022}, enjoins researchers to rigorously and systematically explore the consequences of orthodoxy---to subject it to strict scrutiny.\footnote{Indeed, as an anonymous reviewer pointed out to us (and as is also illustrated by Wheeler's research strategy), both components often went hand in hand in historical practice. In general, our recommended double strategy fits, and is best seen within the framework of, \textcite{Chang2012}'s ``active normative epistemic pluralism". } The hope is thereby to ``test $\Lambda$CDM---our ‘standard model’ of cosmology---to destruction'' (\textcite{Efstathiou}). Again the underlying rationale is broadly falsificationist: further pursuing
the standard model promises relatively straightforward falsification, with significant ramifications
for our theoretical background knowledge.

To conclude, we couldn't agree more with \textcite[p.~429]{Amendola_Tsujikawa_2015}'s spirit of excitement and exploration: ``(e)ven if the nature of dark energy will continue to elude us, all the effort  in this direction will not be in vain. It could as well happen that instead of a shorter route to the East we will find a whole new world.''

\printbibliography

\end{document}